\documentclass[conference]{IEEEtran}
\IEEEoverridecommandlockouts
\usepackage{cite}
\usepackage{amsmath,amssymb,amsfonts}
\usepackage{algorithmic}
\usepackage{graphicx}
\usepackage{textcomp}
\usepackage{xcolor}

\usepackage[T1]{fontenc}
\usepackage[shortlabels]{enumitem} %
\usepackage[autostyle=false, style=english]{csquotes}
\MakeOuterQuote{"}

\usepackage[normalem]{ulem}
\useunder{\uline}{\ul}{}
\usepackage{tabularx}
\usepackage{array}
\usepackage{booktabs}
\usepackage{multirow}
\usepackage{progressbar}
\usepackage{balance}
\usepackage{hyperref}

\usepackage{tikz}
\usepackage{textcomp}

\newcommand{\altparagraph}[1]{%
  \smallskip
  \textbf{#1}
  \noindent
}

\newcommand{\paper}[1]{{\sl P#1}}

\def\BibTeX{{\rm B\kern-.05em{\sc i\kern-.025em b}\kern-.08em
    T\kern-.1667em\lower.7ex\hbox{E}\kern-.125emX}}

\newcommand\copyrighttext{
  \footnotesize \textcopyright~2024 IEEE. Personal use of this material is permitted. Permission from IEEE must be obtained for all other uses, in any current or future media, including reprinting/republishing this material for advertising or promotional purposes, creating new collective works, for resale or redistribution to servers or lists, or reuse of any copyrighted component of this work in other works. DOI: \href{http://dx.doi.org/10.1109/cloudcom62794.2024.00034 }{10.1109/CloudCom62794.2024.00034}.}
\newcommand\copyrightnotice{
\begin{tikzpicture}[remember picture,overlay]
\node[anchor=south,yshift=10pt] at (current page.south) {\fbox{\parbox{\dimexpr\textwidth-\fboxsep-\fboxrule\relax}{\copyrighttext}}};
\end{tikzpicture}
}

\begin{document}

\title{Sampling in Cloud Benchmarking: A Critical Review and Methodological Guidelines}

\author{
    \IEEEauthorblockN{Saman Akbari\IEEEauthorrefmark{1}, Manfred Hauswirth\IEEEauthorrefmark{1}\IEEEauthorrefmark{2}}
    \IEEEauthorblockA{\IEEEauthorrefmark{1}Technische Universität Berlin, Open Distributed Systems, Berlin, Germany}
    \IEEEauthorblockA{\IEEEauthorrefmark{2}Fraunhofer Institute for Open Communication Systems (FOKUS), Berlin, Germany \\
                      Email: \texttt{\{akbari, manfred.hauswirth\}@tu-berlin.de}}
}

\maketitle
\copyrightnotice

\begin{abstract}

Cloud benchmarks suffer from performance fluctuations caused by resource contention, network latency, hardware heterogeneity, and other factors along with decisions taken in the benchmark design. In particular, the sampling strategy of benchmark designers can significantly influence benchmark results. Despite this well-known fact, no systematic approach has been devised so far to make sampling results comparable and guide benchmark designers in choosing their sampling strategy for use within benchmarks.
To identify systematic problems, we critically review sampling in recent cloud computing research.
Our analysis identifies concerning trends: (i) a high prevalence of non-probability sampling, (ii) over-reliance on a single benchmark, and (iii) restricted access to samples. To address these issues and increase transparency in sampling, we propose methodological guidelines for researchers and reviewers. We hope that our work contributes to improving the generalizability, reproducibility, and reliability of research results.
\end{abstract}


\section{Introduction}
\label{sec:introduction}
Most empirical research requires \textit{sampling}, which is the selection of a specific number of (data) items, i.e., samples, from a larger population~\cite{berndtSamplingMethods2020, baltesSamplingSoftwareEngineering2022, clark2022sample, etikanComparisonConvenienceSampling2016}. In cloud benchmarking, sampling involves choosing and executing benchmarks to represent the performance characteristics of cloud services, platforms, or applications while still providing an accurate view of the system under test, because evaluating the complete set of all possible benchmarks is simply not possible due to time and resource constraints.
Sampling in cloud benchmarking is challenging due to the unpredictable performance fluctuations in cloud environments~\cite{laaberSoftwareMicrobenchmarkingCloud2019, leitnerPatternsChaosStudy2016, schirmerNightShiftUnderstanding2023}, caused by resource contention, network latency, hardware heterogeneity, and other external factors which affect the validity of such benchmarks.

In order to better understand how researchers select and execute benchmarks, we critically review sampling in recent cloud computing research.
Surprisingly, there is little knowledge on the rationale for the selection and execution of benchmarks. Furthermore, we find a lack of established sampling guidelines for researchers or reviewers, suggesting that at the moment this is done in an ad-hoc fashion. In the course of this paper, we back up this rather strong claim. To the best of our knowledge, this paper presents the first study that reviews sampling in cloud benchmarking.
Previous work in cloud benchmarking focuses on workload or benchmark design~\cite{rajputEdgefaasbenchBenchmarkingEdge2022, baurleCombFlexibleApplicationoriented2022, wenCharacterizingCommodityServerless2023}, whereas there is little emphasis on \textit{how} researchers select and execute benchmarks, which has a tremendous impact on benchmark results. Meanwhile, current research in sampling primarily addresses its theoretical aspects~\cite{berndtSamplingMethods2020, etikanComparisonConvenienceSampling2016}.

In this paper, we contribute the following: We examine sampling practices in recent cloud computing research and analyze four aspects---sampling methods, sample origin, sample size, and sample availability. Our analysis identifies three concerning trends: almost all samples are non-probabilistic, many studies rely heavily on a single benchmark, and access to most samples is restricted.
To address the identified issues, we propose sampling guidelines for researchers and reviewers.

The paper is structured as follows: Section~\ref{sec:background} outlines the different sampling methods in recent research and elaborates on the challenges of sampling in cloud benchmarking. Section~\ref{sec:study_design} formalizes our critical review of sampling in recent cloud computing research. Section~\ref{sec:study_results} describes the results of our study, whereas in Section~\ref{sec:discussion}, we discuss the results obtained. We provide sampling guidelines for researchers and reviewers in cloud benchmarking in Section~\ref{sec:guidelines} and Section~\ref{sec:related_work} positions our work with respect to related work on sampling and related guidelines. Finally, Section~\ref{sec:conclusion} concludes our paper and proposes future work.

\section{Background}
\label{sec:background}
We start by giving an overview of different sampling methods in quantitative research and then elaborate on the challenges of sampling specifically in the field of cloud benchmarking.

\begin{figure*}[!th]
    \centering
    \includegraphics[width=\textwidth]{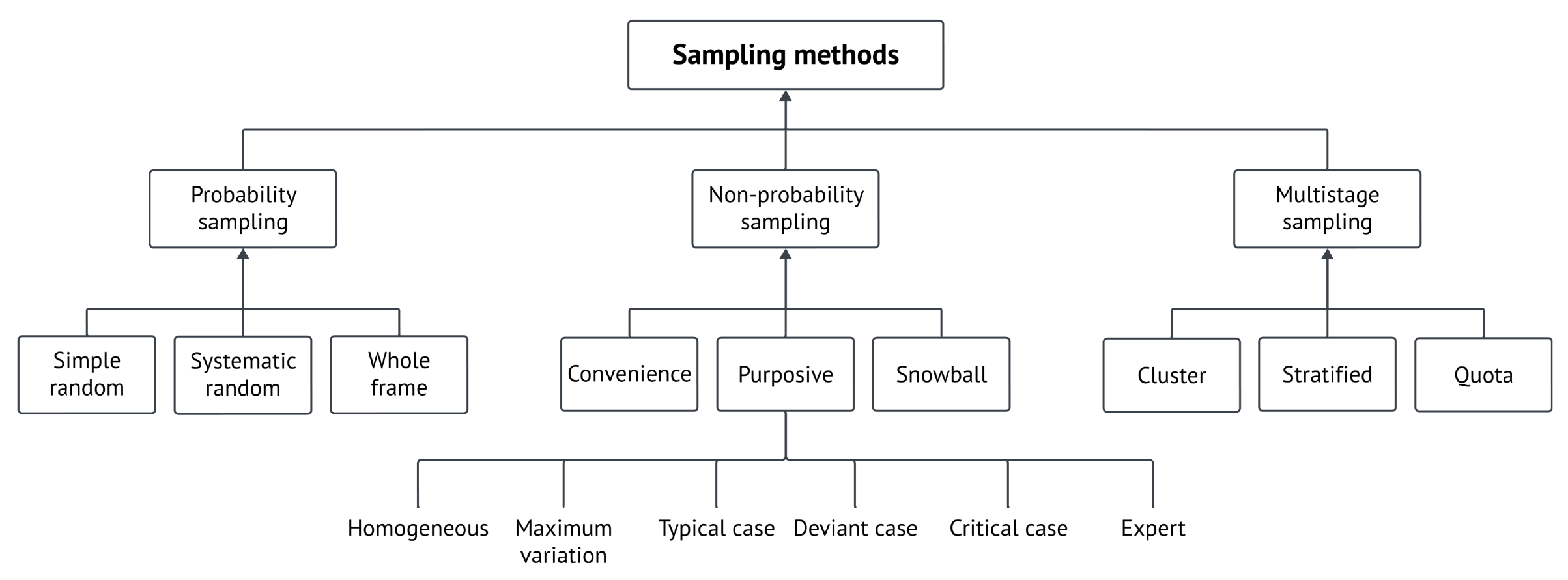}
    \caption{Sampling methods in quantitative research.}
    \label{fig:sampling_methods}
\end{figure*}

\subsection{Sampling Methods}
Typically, drawing conclusions about a population relies on examining a sample. Figure~\ref{fig:sampling_methods} shows the existing sampling methods in quantitative research, which we briefly introduce.

\altparagraph{Probability sampling}
requires that each item in the population has a known and non-zero chance of selection~\cite{clark2022sample}. Common methods include \textit{simple random sampling}, where all items in the population have an equal probability of selection; \textit{systematic random sampling}, which selects items at fixed intervals after a random starting item; and \textit{whole frame sampling}, which selects the entire population~\cite{baltesSamplingSoftwareEngineering2022}. Probability sampling aims for unbiased selection, but requires a complete population list, which is often unavailable. Additionally, systematic random sampling is susceptible to periodic patterns in data.

\altparagraph{Non-probability sampling}
includes methods without randomization, i.e., \textit{convenience sampling}, which selects readily available items; \textit{purposive sampling}, which selects items based on specific criteria~\cite{etikanComparisonConvenienceSampling2016}; and \textit{snowball sampling}, which selects items based on previously selected items. Although non-probability sampling allows researchers to target specific items, it introduces a significant risk of bias~\cite{baltesSamplingSoftwareEngineering2022, berndtSamplingMethods2020}.

\altparagraph{Multistage sampling}
combines sampling different methods. Common techniques include \textit{cluster sampling}, which randomly selects items from subgroups that are similar externally but diverse internally; \textit{stratified sampling}, which randomly selects items from subgroups that are different externally but similar internally; and \textit{quota sampling}, which purposefully selects items from the population to meet defined quotas~\cite{baltesSamplingSoftwareEngineering2022, berndtSamplingMethods2020}. Multistage sampling can offer a balance between the advantages of probability and non-probability methods, but typically requires good knowledge of the underlying population.

\subsection{Challenges of Sampling in Cloud Benchmarking}
The complex nature of cloud environments presents significant challenges for obtaining reliable and accurate benchmark results, as we explain in the following.

\altparagraph{Lack of standardization:}
When evaluating performance, researchers face the task of selecting appropriate benchmarks. This selection process often occurs in an ad-hoc fashion, with little systematic reasoning (see Section~\ref{sec:study_results}). Our personal experience shows that researchers frequently rely on familiarity or convenience when choosing benchmarks. Specifically in cloud computing, the low adoption of benchmark suites exacerbates this issue, due to the rapid evolution of cloud technologies, the diversity of cloud services, and the relative youth of the field. Although benchmark suites exist for cloud computing, such as the SPEC Cloud IaaS benchmark \cite{specCloud2018}, their adoption in the literature remains limited, compared to other fields such as online transaction processing, which have well-defined benchmarks like TPC-C. This motivates our analysis on the \textit{selection of benchmarks} in recent cloud computing literature.

\altparagraph{Performance fluctuations on cloud platforms:}
Cloud platforms experience fluctuations in performance due to factors like resource contention, network latency, and hardware heterogeneity~\cite{laaberSoftwareMicrobenchmarkingCloud2019, leitnerPatternsChaosStudy2016, schirmerNightShiftUnderstanding2023}. This motivates our analysis on the \textit{execution of benchmarks} in recent cloud computing literature.

\altparagraph{Cost and resource constraints:}
Conducting comprehensive cloud benchmarks can be expensive and resource-intensive, especially when aiming for statistically significant sample sizes. Although cost and resource constraints are common challenges in many research fields, they are particularly acute in cloud benchmarking. The pay-as-you-go model of cloud services means that every single experiment directly impacts the research budget, which can lead researchers to make trade-offs in the breadth or depth of their studies.

\section{Study Design}
\label{sec:study_design}
\begin{figure*}[!th]
    \centering
    \includegraphics[width=\textwidth]{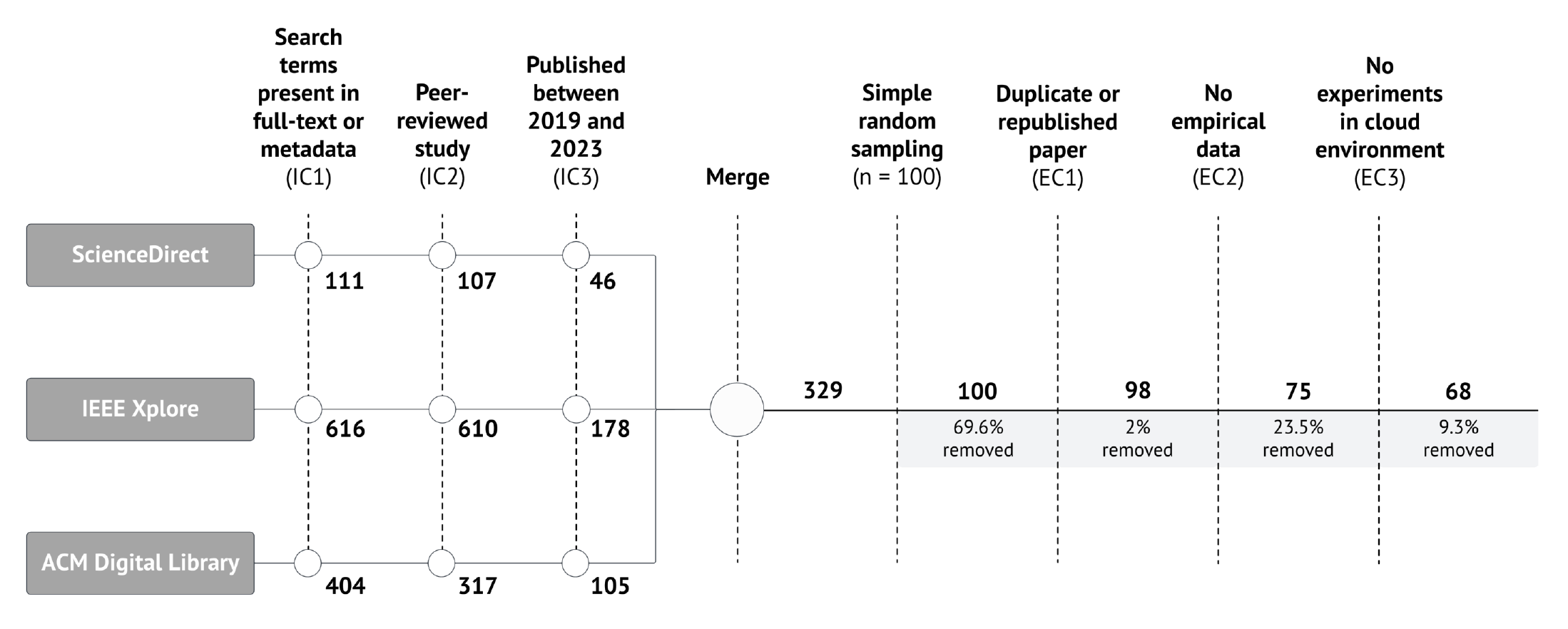}
    \caption{Literature sampling in our critical review.}
    \label{fig:selection_strategy}
\end{figure*}

To address the aforementioned challenge of obtaining reliable and accurate benchmark results in cloud benchmarking, we examine sampling in recent cloud computing research. Our methodology adheres to the critical review guidelines established by Ralph and Baltes~\cite{ralphPavingWayMature2022}. We developed a structured method for identifying and selecting relevant literature, which we depict in Figure~\ref{fig:selection_strategy} and detail in the following.

\subsection{Literature Databases}
\label{sec:literature_databases}
We gathered the literature for our critical review from three reputable literature databases. Our search encompassed the ACM Digital Library, IEEE Xplore, and ScienceDirect.

\subsection{Search String}
\label{sec:search_string}
We now define the search string that we used to query the literature databases. To connect different subjects, we used a logical \texttt{\textbf{AND}}-operator and applied an \texttt{\textbf{OR}}-operator for relevant synonyms. Our analysis focused on performance benchmarks (\texttt{performance \textbf{OR} bench*}) in the cloud domain (\texttt{cloud}), with an emphasis on empirical data from experiments (\texttt{experiment}). Additionally, we integrated two common benchmark terminologies, i.e., \texttt{workload} and \texttt{(metric \textbf{OR} measure)}. The resulting search string that we used is:

\begin{center}
    \texttt{title(performance \textbf{OR} bench*) \textbf{AND} title(cloud) \textbf{AND} experiment \textbf{AND} workload \textbf{AND} (metric \textbf{OR} measure)}
\end{center}

We applied the most important terms to document titles to limit the number of false positives where the terms \texttt{(performance \textbf{OR} bench*)} and \texttt{(cloud)} are merely incidental and not indicative of experiments in a cloud environment. The other terms applied to full-text and metadata. A possible variation of our search string is to include double quotes for exact matching of search terms, but we found that this excludes relevant papers, e.g., due to the use of plurals.

\subsection{Selection Strategy}
\label{sec:selection_strategy}
Our search terms must be present in full text or metadata (\textit{IC1}). Furthermore, we only included peer-reviewed papers, excluding sources such as book chapters, magazines, and early access articles (\textit{IC2}). As the field of cloud computing rapidly evolves, we limited our selection to literature within the last 5 years (2019 -- 2023) (\textit{IC3}).

We then applied simple random sampling with a sample size of 100 papers, because of the substantial volume of available literature in the domain of cloud computing.
We also prescreened the selected papers to exclude papers that could compromise the integrity of our study, and 2 duplicates or republished papers (\textit{EC1}), 23 papers with no empirical data (\textit{EC2}), and 7 papers that do not conduct experiments in a cloud (\textit{EC3}).

Our final sample comprised 68 papers (4 from workshops, 47 from conferences, and 17 from journals) from a total of 115 individual studies. We assigned unique identifiers (\paper{1} to \paper{68}) to each paper, ordered by publication title, which we use to reference the papers in the following sections. 
Our sample is openly accessible (see Section~\ref{sec:open_data}).

\section{Study Results}
\label{sec:study_results}
In this section, we present our results on sampling methods, show sample origins and sample sizes and finally analyze the availability of samples.

\subsection{Sampling Method}
\label{sec:sampling_method_results}
Table~\ref{tab:sampling_methods_results} depicts our results on sampling methods in cloud computing research.
In benchmark selection, purposive sampling is the most common choice, where 57 studies reason that their samples are typical cases. Common quality attributes are \textit{"real-world"} (\paper{26}), \textit{"representative"} (\paper{13}) and \textit{"standard"} (\paper{1}). The second most common sampling method is convenience sampling with 28 studies, where no clear sampling rationale is evident (e.g., \paper{6}), or the authors state availability, such as \textit{"free"} (\paper{63}) and \textit{"simple"} (\paper{27}). In purposive sampling, authors also reason about critical cases in 14 studies (e.g., \textit{"important use case"}), maximum variation in 12 studies (e.g., \textit{"extensive," "diverse," "comprehensive"}), selection based on expertise in 2 studies (e.g., \textit{"experience"}), and homogeneity of benchmarks in 1 study. Additionally, a single study uses simple random sampling (\paper{41}), where they subsample a previous study in the same paper. We did not observe multistage sampling.

A similar pattern is evident in benchmark execution, where most studies use non-probability sampling. Specifically, 57 studies use convenience sampling. Furthermore, purposive sampling is common, especially maximum variation in 12 studies (e.g., varying call frequencies), typical case sampling in 7 studies (e.g., \textit{"single representative run"} (\paper{2}), \textit{"real-world clients"} (\paper{44})), and homogeneous sampling in 7 studies (e.g., load testing). Rarely, researchers also use sophisticated benchmark execution strategies, such as expert sampling in 2 studies (e.g., measurements after a custom-defined warm-up period), or critical case sampling in 2 studies to answer a specific research question. In contrast to benchmark selection, researchers also use probability sampling, namely systematic random sampling in 17 studies (e.g., \textit{"every 4 hours"}), whole frame sampling in 10 studies (e.g., profiling), and simple random sampling in 1 study. We did not observe multistage sampling.

\begin{table*}[!th]
    \centering
    \caption{Sampling methods in cloud computing research ($n = 115$ studies).}
    \label{tab:sampling_methods_results}
    \begin{tabular}{lllcrllcr}
        \toprule
        \multirow{2}{*}{Sampling Method} & & \multicolumn{3}{c}{\textit{Benchmark Selection}} & & \multicolumn{3}{c}{\textit{Benchmark Execution}} \\ \cmidrule(lr){3-5} \cmidrule(lr){7-9}
        
        & \hspace{1cm} & & Count & \multicolumn{1}{c}{Percentage} & \hspace{0.8cm} & & Count & \multicolumn{1}{c}{Percentage} \\
        
        \midrule
        \textbf{Probability}        & & &  &  &  &  \\ \addlinespace
        \quad Systematic random     & & & 0 & 0\% \progressbar[filledcolor=black!50]{0}
                                    & & & 17 & 14.8\% \progressbar[filledcolor=black!50]{0.148} \\ \addlinespace
        \quad Whole frame           & & & 0 & 0\% \progressbar[filledcolor=black!50]{0}
                                    & & & 10 & 8.7\% \progressbar[filledcolor=black!50]{0.087} \\ \addlinespace
        \quad Simple random         & & & 1 & 0.9\% \progressbar[filledcolor=black!50]{0.009}
                                    & & & 1 & 0.9\% \progressbar[filledcolor=black!50]{0.009} \\ \addlinespace
        \midrule
        \textbf{Non-Probability}    & & &  &  &  &  \\ \addlinespace
        \quad Convenience           & & & 28 & 24.3\% \progressbar[filledcolor=black!50]{0.243}
                                    & & & 57 & 49.6\% \progressbar[filledcolor=black!50]{0.496} \\ \addlinespace
        \quad Purposive             & & &  &  &  &  \\ \addlinespace
        \qquad i) Maximum variation & & & 12 & 10.4\% \progressbar[filledcolor=black!50]{0.104}
                                    & & & 12 & 10.4\% \progressbar[filledcolor=black!50]{0.104} \\ \addlinespace
        \qquad ii) Typical case     & & & 57 & 49.6\% \progressbar[filledcolor=black!50]{0.496}
                                    & & & 7 & 6.1\% \progressbar[filledcolor=black!50]{0.061} \\ \addlinespace
        \qquad iii) Critical case   & & & 14 & 12.2\% \progressbar[filledcolor=black!50]{0.122}
                                    & & & 2 & 1.7\% \progressbar[filledcolor=black!50]{0.017} \\ \addlinespace
        \qquad iv) Expert           & & & 2 & 1.7\% \progressbar[filledcolor=black!50]{0.017}
                                    & & & 2 & 1.7\% \progressbar[filledcolor=black!50]{0.017} \\ \addlinespace
        \qquad v) Deviant case      & & & 0 & 0\% \progressbar[filledcolor=black!50]{0}
                                    & & & 0 & 0\% \progressbar[filledcolor=black!50]{0} \\ \addlinespace
        \qquad vi) Homogeneous      & & & 1 & 0.9\% \progressbar[filledcolor=black!50]{0.009}
                                    & & & 7 & 6.1\% \progressbar[filledcolor=black!50]{0.061} \\ \addlinespace
        \quad Snowball              & & & 0 & 0\% \progressbar[filledcolor=black!50]{0}
                                    & & & 0 & 0\% \progressbar[filledcolor=black!50]{0} \\ \addlinespace
        \midrule
        \textbf{Multistage}         & & &  &  &  &  \\ \addlinespace
        \quad Stratified            & & & 0 & 0\% \progressbar[filledcolor=black!50]{0}
                                    & & & 0 & 0\% \progressbar[filledcolor=black!50]{0} \\ \addlinespace
        \quad Quota                 & & & 0 & 0\% \progressbar[filledcolor=black!50]{0}
                                    & & & 0 & 0\% \progressbar[filledcolor=black!50]{0} \\ \addlinespace
        \quad Cluster               & & & 0 & 0\% \progressbar[filledcolor=black!50]{0}
                                    & & & 0 & 0\% \progressbar[filledcolor=black!50]{0} \\ \addlinespace
        \bottomrule
    \end{tabular}
\end{table*}

\subsection{Sample Origin}
\label{sec:sample_origin_results}
Table~\ref{tab:sample_origin_results} shows the origin of the benchmarks in cloud computing research. Most often, researchers use existing benchmarks from open-source platforms (34), such as GitHub. Additionally, researchers create custom benchmarks specifically for their study (28), and use benchmarks based on standardized methodologies (25), i.e., benchmarks that allow reliable and portable performance comparisons across different systems by having well-defined workload characteristics, procedures, and metrics. In some cases, the sample origin is unclear (15). Researchers also use generated benchmarks (8) and subsamples (5) from previous studies in the same paper.

\begin{table}[!th]
    \centering
    \caption{Sample origin of benchmarks in cloud computing research ($n = 115$ studies).}
    \label{tab:sample_origin_results}
\begin{tabular}{lrr}
    \toprule
    Sample Origin                   & Count     & \multicolumn{1}{c}{Percentage} \\ \midrule
    \textbf{Open source}            & 34        & 29.6\% \progressbar[filledcolor=black!50]{0.296}     \\
    \textbf{Custom}                 & 29        & 25.2\% \progressbar[filledcolor=black!50]{0.252}   \\
    \textbf{Standardized}           & 25        & 21.7\% \progressbar[filledcolor=black!50]{0.217}   \\
    \textbf{Unclear}                & 15        & 13\% \progressbar[filledcolor=black!50]{0.13}     \\
    \textbf{Generated}              & 8         & 7\% \progressbar[filledcolor=black!50]{0.07}    \\
    \textbf{Subsample}              & 4         & 3.5\% \progressbar[filledcolor=black!50]{0.035}     \\
    \bottomrule
\end{tabular}
\end{table}

\subsection{Sample Size}
\label{sec:sample_size_results}
\begin{figure}[!th]
  \centering
    \includegraphics[width=\columnwidth]{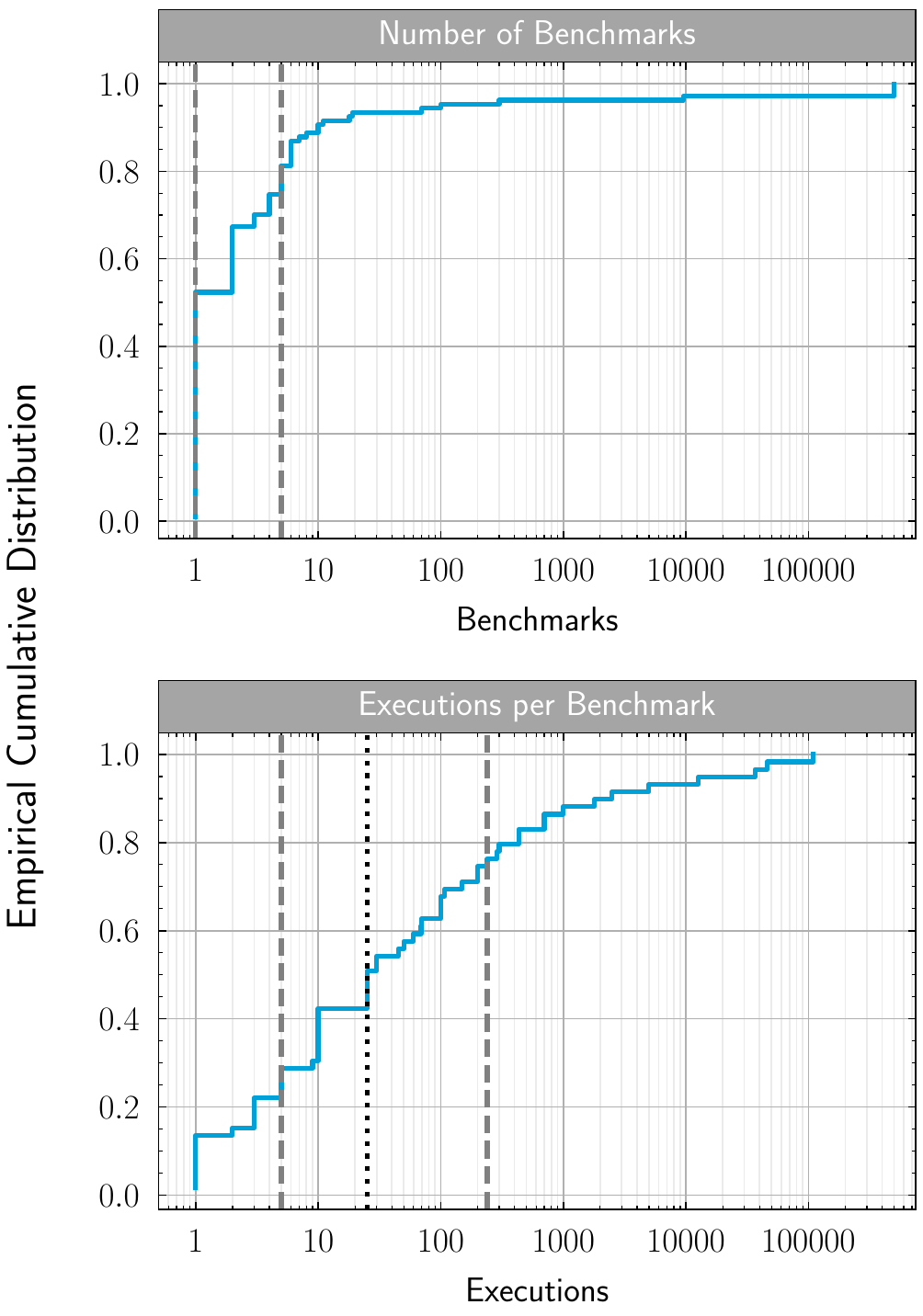}
    \caption{Empirical cumulative distribution of benchmarks ($n = 107$ studies) and executions ($n = 59$ studies). Left dashed line: p25, dotted line: median, right dashed line: p75.}
    \label{fig:ecdf}
\end{figure}

Figure~\ref{fig:ecdf} depicts the empirical cumulative distribution functions (ECDF) on a logarithmic scale for the number of different benchmarks per study as well as for the executions per benchmark.
The benchmark selection is skewed towards low numbers of benchmarks, with the majority (52.3\%) of studies using only 1 benchmark. Furthermore, 90.7\% of studies use 10 benchmarks or fewer. The long tail in the distribution indicates that few studies employ many benchmarks (e.g., $>$70 benchmarks), which are mostly generated. In 8 of 115 studies (7\%), the number of benchmarks is either unspecified or unclear.

For the executions per benchmark, the specifics are unclear or unspecified in 56 of 115 studies (48.9\%). The ECDF shows a more varied distribution compared to the number of benchmarks. Some studies perform as few as a single execution (13.6\%), whereas there are outliers with many executions reaching up to 110,000 executions in a load testing study (\paper{3}). The 25th percentile is 5 executions, the median is 25 executions, and the 75th percentile is 220 executions.

\subsection{Sample Availability}
\label{sec:sample_availability_results}
59 of 115 studies provide implicit access to benchmarks due to their sample origin being open-source or standardized benchmarks. In the remaining 56 studies, only 12 studies provide access to benchmarks. Therefore, in total, 71 of 115 studies (61.7\%) provide access to benchmarks.

5 studies provide partial access to benchmarks: (i) 1 study references broken links, (ii) 2 studies reference common algorithms without providing or indicating specific implementations, and (iii) 2 studies reference benchmarks that require paid access.

Only 17 of 115 studies (14.8\%) publish their sampled data, whereas 1 study provides partial access due to broken links.

\section{Discussion}
\label{sec:discussion}

In this section, we discuss our results on sampling methods, sample origins, sample sizes, and sample availability. Then, we elaborate on limitations when interpreting our results and finally, discuss threats to the validity of our work.

\subsection{Sampling Method}
\label{sec:discussion_sampling_method}
Probably the biggest limitation in cloud benchmarking relates to the concept of a sampling frame. Probability sampling requires a well-defined sampling frame, which represents the population of interest~\cite{baltesSamplingSoftwareEngineering2022}. However, it is impossible to define a sampling frame, which contains all possible benchmarks. Therefore, it would be unfair to criticize researchers for using non-probability sampling.

Nevertheless, non-probability sampling limits the generalizability of research findings~\cite{berndtSamplingMethods2020, baltesSamplingSoftwareEngineering2022, etikanComparisonConvenienceSampling2016}, which is a concern.
Surprisingly, in benchmark execution, where the sampling frame, i.e., time, is a readily defined concept unlike benchmark selection, convenience and purposive sampling are also prevalent. Only 28 of 115 studies use probability sampling.
If researchers execute benchmarks during off-peak hours when cloud infrastructures are underutilized, the measured performance may not reflect real-world scenarios accurately~\cite{laaberSoftwareMicrobenchmarkingCloud2019, leitnerPatternsChaosStudy2016, schirmerNightShiftUnderstanding2023}. Conversely, if researchers execute benchmarks only during peak hours, they may not capture a system's performance under typical usage conditions. Therefore, non-probability sampling in benchmark execution also impacts the generalizability of the results obtained.

\subsection{Sample Origin}
\label{sec:discussion_sample_origin}
15 of 115 studies lack clear information regarding the origin of benchmarks, which impairs the assessment of potential bias. Knowing the source of benchmarks can shed light on potential conflicts of interest or limitations in the benchmarks themselves.
However, when authors specify benchmark origins, they often originate from open source. Custom benchmarks are also common to address specific research needs.
Although 25 of 115 studies use a standardized benchmark, it would be beneficial that this number would be higher, because it would make it easier to compare the findings of various studies. The relatively low adoption of standardized benchmarks can be attributed partially to the youth of cloud computing, in contrast to established fields such as online transaction processing, which have well-defined benchmarks like TPC-C.

\subsection{Sample Size}
\label{sec:discussion_sample_size}
A surprising finding is that approximately half of the studies rely on a single benchmark for evaluation,
which poses risks to the generalizability of results if the benchmark does not capture diverse workload types. Researchers might be compelled to focus on a single benchmark due to the time and costs of running and analyzing multiple benchmarks, but this is a problem that should be addressed. By relying on a single benchmark, results potentially lead to biased or incomplete insights.

The number of executions per benchmark varies, ranging from a single execution to large numbers, e.g., 110,000 in \paper{3}. Although repeated executions are crucial for statistical robustness, they do not eliminate potential bias in the selected benchmark or methodology~\cite{baltesSamplingSoftwareEngineering2022}. In other words, no matter how many times a biased benchmark is repeated, inherent bias remains unaddressed. Furthermore, the sample size depends on the specific type of benchmark used. For micro-benchmarks designed to assess specific, isolated aspects of performance, many executions might be feasible. However, for macro-benchmarks or application-level benchmarks, which simulate complex workloads, the time and cost associated with many executions might be prohibitive.

\subsection{Sample Availability}
\label{sec:discussion_sample_availability}
The limited availability of both benchmarks and sampled data significantly impairs the replicability of research findings.
59 of 115 studies use open-source or standardized benchmarks, whereas only 12 of the remaining 56 studies publish their benchmarks.
Furthermore, only 18 of 115 studies grant access to their sampled data. Ideally, researchers would share all relevant materials, including the specific benchmarks used \textit{and} raw data generated from experiments. Publishing benchmarks and data requires minimal effort and cost, but significantly improves the credibility, validity, and reproducibility of research.
Furthermore, as 2 studies with broken links show, researchers should preferably use open-access repositories that provide permanent links, such as Zenodo.

\subsection{Limitations}
\label{sec:discussion_limitations}
Our study has a few limitations: It does not consider the configuration of cloud benchmarks in studies, e.g., the number of nodes, the cloud provider, or the geographical region. Given the broad scope of our study on cloud computing, an analysis of potential configurations is not particularly informative, however.

Moreover, we do not differentiate between various qualities of service (QoS) in cloud computing, such as performance, availability, security, elastic scalability, and data consistency~\cite{bermbachCloudServiceBenchmarking2017}. Sampling is certainly influenced by the specific QoS in question, for example security testing requires a different sampling strategy compared to elastic scalability testing. However, a comprehensive analysis of all potential QoS considerations is beyond the scope of this paper.

In benchmark execution, we primarily address the timing of execution. We do not consider the sampling of benchmark inputs. Such an analysis is of minimal value in our study due to the heterogeneity of benchmarks.

\subsection{Threats to Validity}
\label{sec:threats_to_validity}

This section discusses the potential threats to the validity of our results in the critical review on sampling in cloud benchmarking, and the actions we have taken to mitigate them.

\altparagraph{External validity:} 
Our findings may not be generalizable to the entire literature on cloud computing. We may have only captured a disproportionate number of studies with inadequate sampling practices. To address this concern, we implemented the following actions: (i) We restricted our analysis to peer-reviewed papers, e.g., excluding book chapters, (ii) we focused on well-established publishers known for rigorous review processes (ACM, IEEE, and ScienceDirect), and (iii) we limited our scope to papers published within the past five years to capture recent research and practices in cloud benchmarking.

\altparagraph{Construct validity:}
The accuracy of our findings relies on correctly identifying sampling strategies in the studies that we reviewed. Textual descriptions of sampling strategies can be ambiguous, and authors may not always provide clear explanations. For example, we sometimes derived data from tables or figures within the studies. These ambiguities can introduce threats to construct validity, requiring expertise to accurately interpret sampling strategies. We mitigated these threats through the following actions: (i) We allocated sufficient time for each paper review (approximately 45 minutes) to ensure a thorough analysis, (ii) we implemented a structured review process that involved extracting verbatim quotes, and (iii) we documented and commented on any cases where the sampling strategy was unclear for traceability (see Section~\ref{sec:open_data}).

\altparagraph{Internal validity:}
Concerns about internal validity are minimal in our study. There is a clear causal relationship between textual descriptions of sampling and our reported results on sampling methods, sample size, sample origin, and sample availability.

\altparagraph{Replicability:}
To facilitate the replication of our findings and independent verification, we published a comprehensive replication package described in Section~\ref{sec:open_data}.

\section{Methodological Guidelines}
\label{sec:guidelines}
The results in Section~\ref{sec:study_results} show a pervasive lack of transparency in cloud computing studies when describing sampling strategies, which hinders the accurate interpretation of results. To mitigate this issue, we propose four sampling guidelines for researchers and reviewers, grounded in the \textit{ACM SIGSOFT Empirical Standards} for metascience~\cite{ralphACMSIGSOFTEmpirical2021}. Our proposed guidelines are as follows:

\leavevmode\begin{enumerate}[label=G\arabic*:, itemsep=1em]
    \item \textbf{Clear sampling rationale}\\
    State the rationale for the selection and execution of your benchmarks. Explain how your chosen benchmarks address specific performance aspects (e.g., scalability, CPU, memory). Detail your execution strategy and consider the performance fluctuations on cloud platforms. Having a clear sampling rationale helps the reader interpret your methodology and results.
    
    \item \textbf{Transparent sampling description}\\    
    Provide enough detail about your sample origin and sampling procedure, so that an informed reader can replicate your sampling. Transparent sampling descriptions allow better replication of your study.

    \item \textbf{Acknowledgment of potential sampling bias}\\
    Proactively acknowledge and discuss potential sources of bias inherent in your sampling strategy to show that you are aware of limitations and increase the credibility of your work.
    
    \item \textbf{Open access}\\
    Publish all artifacts, including the code of custom-made benchmarks, raw performance data, and any scripts for deployment or execution. Use permanent repositories like Zenodo to ensure long-term accessibility. Be careful to redact or anonymize possibly confidential data. This guideline addresses the current lack of benchmark and data availability in cloud computing research.
\end{enumerate}

\leavevmode\\ With respect to these methodological guidelines for sampling, we want to highlight some positive examples that we found in our study (in parentheses the guideline which was adhered to):

\begin{itemize}
    \item \paper{11} elaborates that they select eight benchmarks with varying levels of CPU and memory intensity to \textit{"represent a wide range of applications"} (G1).
    \item \paper{16} accurately describes their benchmark in edge-cloud clusters, the number of executions, and the time between executions (G2).
    \item \paper{39} explains that \textit{"addressing the performance variations experienced by cloud applications was not the focus of [their] study"} (G3).
    \item \paper{10} open-sources their custom benchmarks, the source code for execution of the benchmarks, and the sampled data (G4).
\end{itemize}

\begin{figure}[!th]
  \centering
    \includegraphics[width=\columnwidth]{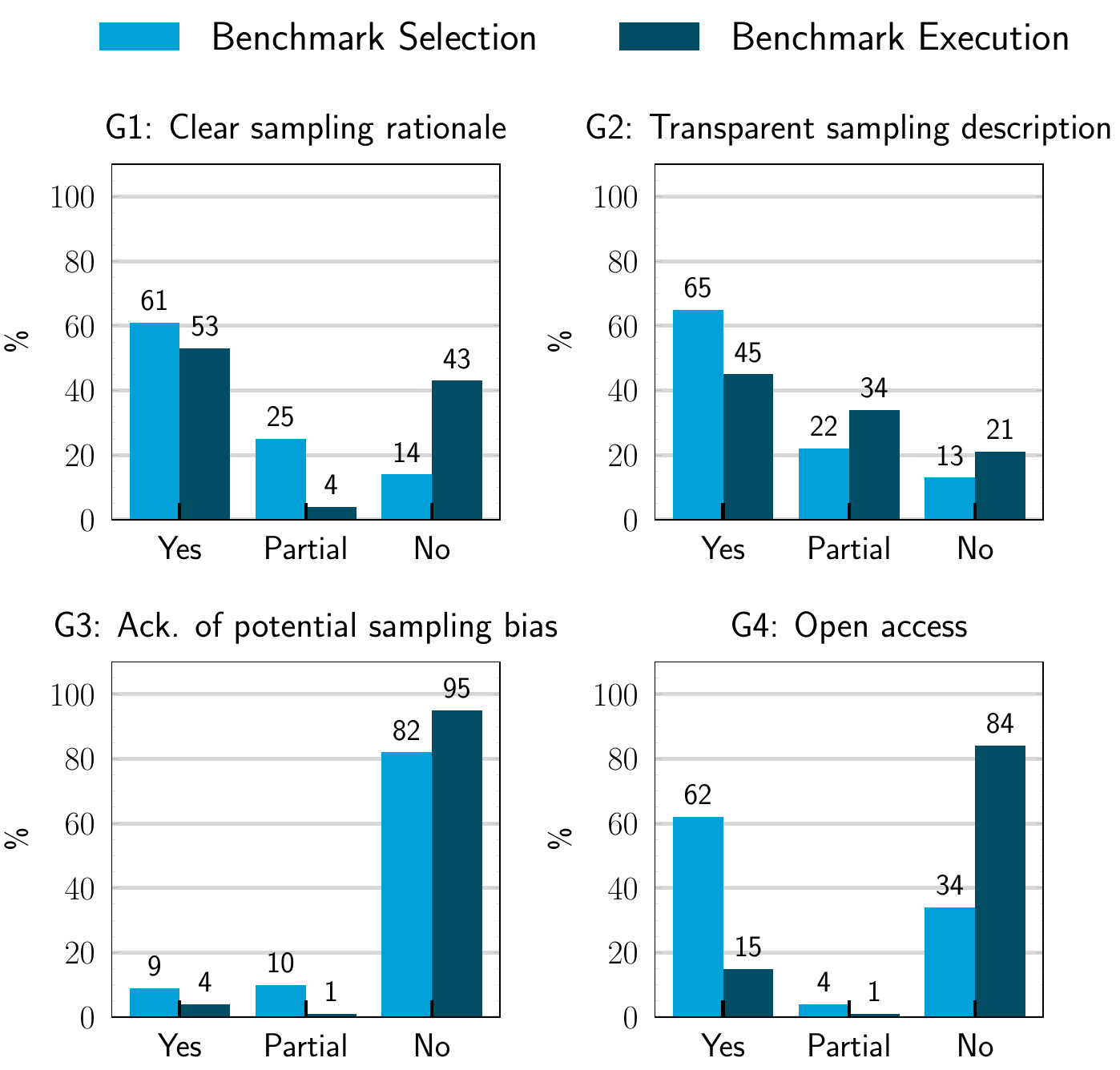}
    \caption{Evaluation of sampling guidelines G1--G4 ($n = 115$ studies).}
    \label{fig:guidelines_evaluation}
\end{figure}

\leavevmode\\ Figure~\ref{fig:guidelines_evaluation} shows to what extent studies in recent cloud computing research follow the methodological guidelines for sampling.

The majority of studies ($>$50\%) have a clear sampling rationale (G1) in both the selection and the execution of benchmarks. In some studies, we can make an educated guess, i.e., authors do not explicitly state any reason, which we record as partial fulfillment (e.g., \paper{28}).
We see a similar pattern in providing a transparent sampling description (G2), where we could most likely replicate the sampling in approximately half of the studies based on their description alone. Some studies omit important details, such as the sample origin, e.g., \paper{4}, \paper{65}, whereas other studies do not provide clear descriptions, e.g., "several micro-services" in \paper{3}.
Most studies struggle with acknowledging potential sample bias (G3). In benchmark selection, 82\% of the studies and in benchmark execution even 95\% of the studies do not mention any threats to validity or limitations.
Lastly, we have open access (G4) to the benchmarks in 62\% of studies, primarily due to open-source, e.g., \paper{54}, and standardized benchmarks, e.g., \paper{11}. In contrast, most studies do not provide open access to their sampled data (84\%). We record paid access to benchmarks, e.g., \paper{48}, ambiguities in the choice of implementation, e.g., \paper{57}, and broken links, e.g., \paper{55}, as partial fulfillment of this guideline.
Overall, we find a lack of sampling transparency in benchmark execution, compared to benchmark selection.

\section{Related Work}
\label{sec:related_work}

There is a substantial body of literature dedicated to sampling. One research direction discusses the theory of sampling. For example, Berndt~\cite{berndtSamplingMethods2020} explains the advantages and disadvantages of different sampling methods. Likewise, Etikan et al.~\cite{etikanComparisonConvenienceSampling2016} conduct a comparative analysis between convenience and purposive sampling. These theoretical discussions informed our discussion of sampling methods in Section~\ref{sec:background}.

Another line of research examines sampling in different research methodologies~\cite{cashSamplingDesignResearch2022}, research fields~\cite{amirThereNoRandom2018, baltesSamplingSoftwareEngineering2022}, or a synthesis of both~\cite{hieblSampleSelectionSystematic2023}. Our work is in the category of sampling within a research field, i.e., cloud benchmarking.

The term \textit{sampling} is not common in cloud benchmarking, but appears under the "disguise" of other terms, such as "selection of workloads"~\cite{muhlbauerAnalysingImpactWorkloads2023}. Research in cloud benchmarking mostly deals with workload or benchmark design, often as benchmark suites~\cite{rajputEdgefaasbenchBenchmarkingEdge2022, baurleCombFlexibleApplicationoriented2022, wenCharacterizingCommodityServerless2023}. However, the rationale behind how researchers select and execute benchmarks remains largely unexplored.
The closest work to our research is that of Baltes and Ralph~\cite{baltesSamplingSoftwareEngineering2022} on sampling in software engineering. Similarly to recent cloud computing research, software engineering (SE) research shows limited use of probability sampling (15.3\% in SE vs. our 0.9\% in cloud computing for benchmark selection), ambiguities in sampling descriptions (24\% vs. 13\%), and a lack of transparency. However, their work does not address sampling in benchmark execution, and the areas of sample size or sample availability.

In the field of benchmarking, Bartz-Beielstein et al.~\cite{bartz-beielsteinBenchmarkingOptimizationBest2020} discuss best practices and formulate guidelines for researchers on benchmark selection. Their recommendations entail the properties of diversity, representativeness, scalability, and known solutions. Together with our proposed sampling guidelines, they achieve synergistic goals: Bartz-Beielstein et al. ensure that benchmarks are technically sound and diverse, while our guidelines ensure that the process of selecting and executing benchmarks is transparent.

To the best of our knowledge, in cloud benchmarking, sampling guidelines have not yet been formulated prior to our work. The closest work is by Papadopoulos et al.~\cite{papadopoulosMethodologicalPrinciplesReproducible2019}, who propose eight principles for reproducible performance evaluation in cloud computing. These principles primarily address designing experiments and reporting results. Interestingly, their open access principle aligns with our G4 guideline, and their emphasis on comprehensive experimental setup description is similar to our guideline on transparent sampling description (G2).

\section{Conclusion \& Future Work}
\label{sec:conclusion}

In this paper, we critically reviewed the state of sampling in cloud benchmarking across 115 recent studies. Our work aims to raise awareness of \textbf{transparency} in sampling.
In our analysis, we found a dominance of non-probability sampling and researchers frequently use convenience and purposive sampling in both the selection and the execution of benchmarks. This sampling method raises concerns about the generalizability of research findings in the corresponding papers. Also, a significant portion of studies rely on a single benchmark. Limited access to both benchmarks and sampled data further impedes replicability.
To address the lack of transparency in sampling, we defined four methodological guidelines for researchers and reviewers.

Future work could collect available cloud benchmarks to obtain comprehensive sampling frames, which would serve as the foundation for probability sampling. Stratification based on factors like cloud service type, or programming language of the implementation is possible. However, maintaining sampling frames presents challenges, such as how new benchmarks will be incorporated, e.g., automated or curated selection. This work most likely requires collaboration and agreement in a larger community of interested researchers.
As a first step, researchers could focus on developing and promoting standardized benchmarks. Just as the TPC-C benchmark is the standard for evaluating online transaction processing systems, cloud computing needs similar benchmarks to improve the comparability and generalizability of research results.

\section{Open Data}
\label{sec:open_data}
We published a replication package~\cite{akbari2024} on Zenodo, which includes (i) all search queries for our critical review along with links to the original search engines, (ii) query results in \BibTeX~format, (iii) the source code for literature sampling and generating our plots, and (iv) structured explanations detailing our literature pre-screening and critical review, including verbatim quotes from the papers describing the sampling strategies employed.

\balance
\bibliographystyle{bibliography/IEEEtran}
\bibliography{bibliography/IEEEabrv,bibliography/IEEEbibliography}

\end{document}